NETWORK SCIENCE

# Resilience and efficiency in transportation networks

Alexander A. Ganin,[1,2] Maksim Kitsak,[3] Dayton Marchese,[2] Jeffrey M. Keisler,[4] Thomas Seager,[5] Igor Linkov[2]*

Urban transportation systems are vulnerable to congestion, accidents, weather, special events, and other costly delays. Whereas typical policy responses prioritize reduction of delays under normal conditions to improve the efficiency of urban road systems, analytic support for investments that improve resilience (defined as system recovery from additional disruptions) is still scarce. In this effort, we represent paved roads as a transportation network by mapping intersections to nodes and road segments between the intersections to links. We built road networks for 40 of the urban areas defined by the U.S. Census Bureau. We developed and calibrated a model to evaluate traffic delays via link loads. The loads may be regarded as traffic-based centrality measures, estimating the number of individuals using corresponding road segments. Efficiency was estimated as the average annual delay per peak-period auto commuter, and modeled results were found to be close to observed data, with the notable exception of New York City. Resilience was estimated as the change in efficiency resulting from roadway disruptions and was found to vary between cities, with increased delays due to a 5% random loss of road linkages ranging from 9.5% in Los Angeles to 56.0% in San Francisco. The results demonstrate that many urban road systems that operate inefficiently under normal conditions are nevertheless resilient to disruption, whereas some more efficient cities are more fragile. The implication is that resilience, not just efficiency, should be considered explicitly in roadway project selection and justify investment opportunities related to disaster and other disruptions.

## INTRODUCTION

Existing roadway design standards emphasize the efficient movement of vehicles through a transportation network (1–4). Efficiency in this context may include identification of the shortest or fastest route (1, 5–7), or the route that minimizes congestion (8). It is the primary criterion on which road networks are modeled and design alternatives are considered (6, 7, 9, 10). The Texas A&M Transportation Institute defines and reports traffic delay in urban areas as the annual delay per auto commuter (11). Other studies define efficiency as delay for the individual driver in terms of time spent moving or stopped (7), or mean travel time between all origin-destination pairs in the network (9). However, as the experience of any motorist in large American cities can attest, conditions beyond the scope of the roadway design, including congestion, accidents, bad weather, construction, and special events (for example, a marathon race), can cause costly delays and frustrating inefficiencies that result in fuel waste, infrastructure deterioration, and increased pollution (12, 13). Evaluating road networks based only on efficiency under normal operating conditions results in little to no information about how the system performs under suboptimal or disrupted conditions.

Infrastructure systems that exhibit adaptive response to stress are typically characterized as resilient (14–21). Given the essential role of transportation in emergency response, provision of essential services, and economic well-being, the resilience of roadway networks has received increasing policy attention. Nonetheless, scholars have yet to converge on a shared understanding of resilience suitable to guide design, operation, and reconstruction of roadway networks. Although resilience in infrastructure systems is characterized as a multidimensional concept (22, 23), in many engineering and civil infrastructure implementations, resilience is defined as the ability of a system to prepare for, absorb, recover from, and adapt to disturbances (16). Specific to transportation, resilience has been defined as "the ability of the system to maintain its demonstrated level of service or to restore itself to that level of service in a specified timeframe" (24). Others describe transportation resilience as simply the ability of a system to minimize operational loss (25) or use the term synonymously with robustness, redundancy, reliability, or vulnerability (26–28).

Current efforts in transportation resilience research have focused on framework development and quantification methods. These efforts include the specification of resilience indicators, such as total traffic delay (24), economic loss (29), post-disaster maximum flow (30), and autonomous system components (31). Practical concerns with this type of resilience evaluation are that it relies on uncertain performance data and often omits indicators that are unquantifiable (19). Other resilience approaches apply traffic network modeling to identify locations for critical buildings (for example, hospitals and fire stations) (32), minimize trip distance for individual passengers (33), and minimize travel time across the system (12). One drawback of existing network resilience methods is that they are data-intensive, often requiring limited information about resources for unusual road system repair (26, 28) or network behavior following a disruptive event (34). Moreover, existing resilience quantification approaches lack calibration and testing across a range of transportation systems. Because many disruptive events, and their associated consequences, are difficult to predict, resilient road systems must be characterized and evaluated by the capacity to adapt to a variety of different stress scenarios. Partly because of these obstacles, joint consideration of efficiency and resilience has yet to be implemented for transportation networks.

Here, we study the interconnections between resilience and efficiency (20) among road transportation networks in 40 major U.S. cities. We develop an urban roadway efficiency model, calibrate it on the basis of the observed data (11) of annual delay per peak-period auto commuter, and apply the model to calculate efficiency in 40 cities.

[1]Department of Systems and Information Engineering, University of Virginia, 151 Engineer's Way, P.O. Box 400747, Charlottesville, VA 22904, USA. [2]Engineer Research and Development Center, U.S. Army Corps of Engineers, 696 Virginia Road, Concord, MA 01742, USA. [3]Department of Physics, Northeastern University, 110 Forsyth Street, Boston, MA 02115, USA. [4]College of Management, University of Massachusetts Boston, 100 Morrissey Boulevard, Boston, MA 02125, USA. [5]School of Sustainable Engineering and the Built Environment, Arizona State University, 781 S Terrace Road, Tempe, AZ 85287, USA.
*Corresponding author. Email: igor.linkov@usace.army.mil

Then, we model traffic response to random roadway disruptions and recalculate expected delays to determine the sensitivity of each city to loss of roadway linkages. The results may reveal important considerations for assessing proposals for improvement of roadway infrastructure that maintain efficiency under stress conditions.

## METHODS

The Methods section appears here to help clarify the subsequent sections. To develop the urban roadway efficiency model, we defined the urban area boundaries, constructed the road networks, and evaluated the population density within cities using the Census Bureau data sets (35, 36) and OpenStreetMap (OSM) data sets (37). We relied on these data to assess commuter patterns, which we used to measure efficiency and resilience of road networks.

Alternative approaches to transportation have been offered and include those based on percolation theory and cascading failures (38–40), human mobility pattern studies (41–43), queueing (44, 45), and the use of historical data to predict traffic. We review these approaches in the Supplementary Materials and note that the main benefit of our model is that it relies solely on readily available public data, rather than on particular data sets that may or may not be practical to obtain for any particular region. The model's algorithmic simplicity allows us to consider spatial topologies of cities in high resolution including tens of thousands of nodes and links. We did not create a more accurate transportation model than the existing ones, but we were able to obtain measurable characteristics of transportation systems (average delays) using our model.

## Geospatial boundaries and population density

To define geospatial boundaries for the transportation infrastructure networks, we used the U.S. Census Bureau geospatial data set (35) for urban areas—densely developed residential, commercial, and other nonresidential areas (46). We approximated the exact urban area polygon with a simplified manually drawn one (Fig. 1A) and included all roadways within 40 km (25 miles) of it in the network. For each of the links, we calculated its length on the basis of the polyline defining the link and assigned a number of lanes $m$ and the FFSs (see the Supplementary Materials).

We next estimated population in vicinity of each intersection $i$ using the Census Tract data (36). To this end, we split the map into Voronoi cells centered at intersections and then evaluated the population of each cell $N_i$ as

$$N_i = \sum_t N_t \frac{\text{Area}(P_t \cap P_i)}{\text{Area}(P_t)} \tag{1}$$

Above, $N_t$ is the population of Census Tract $t$, and $P_i$ and $P_t$ are the polygons of the cell and the tract, respectively (Fig. 1B and table S2).

## Transportation model

We built on the gravity model to generate commuting patterns. The gravity model (47) is a classical model for trip distribution assignment and is extensively adopted in most metropolitan planning and statewide travel demand models in the United States (48–51). Other trip distribution models include, for example, destination choice models (52, 53). However, these models are not as widely used in large scale, because the detailed data required by these models are frequently unavailable (48).

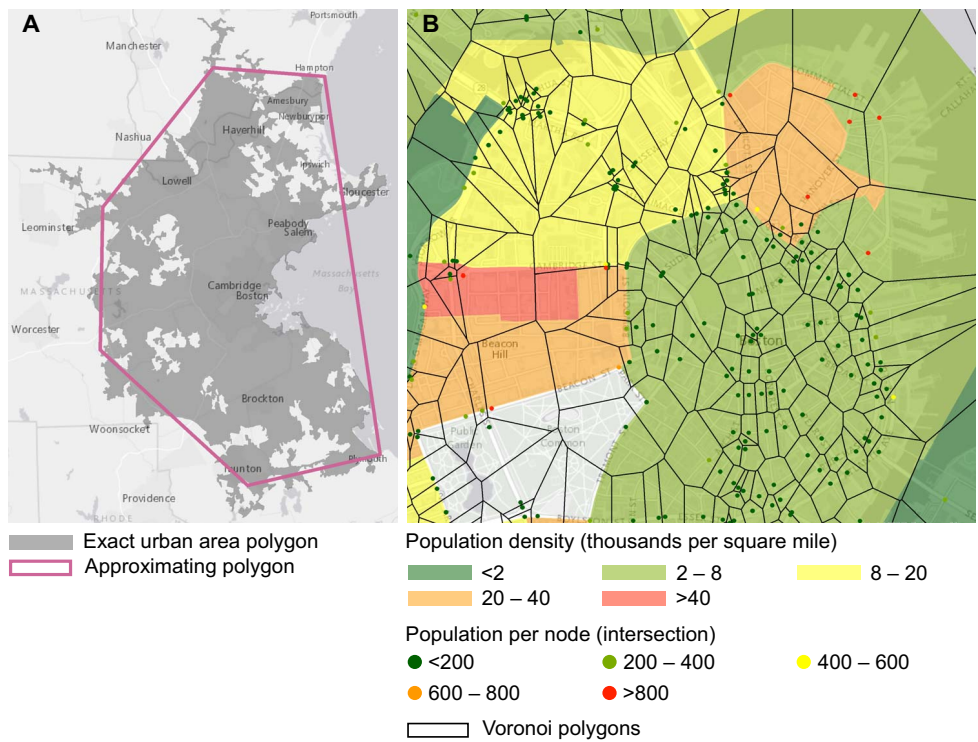

**Fig. 1. Definition of urban areas and assignment of nodes' population.** (**A**) Boston, MA-NH-RI urban area as defined by the U.S. Census Bureau shapefiles (gray background). To simplify the model and the algorithms calculating the distance from network nodes to the city boundary, we approximate each of the urban areas shapefiles with a coarse manually drawn polygon (pink outline). (**B**) Assignment of the number of people departing from each of the network nodes. Population distribution (color polygons; red corresponds to higher population density), Voronoi polygons (black outline), and network nodes (dots) in Downtown Boston.

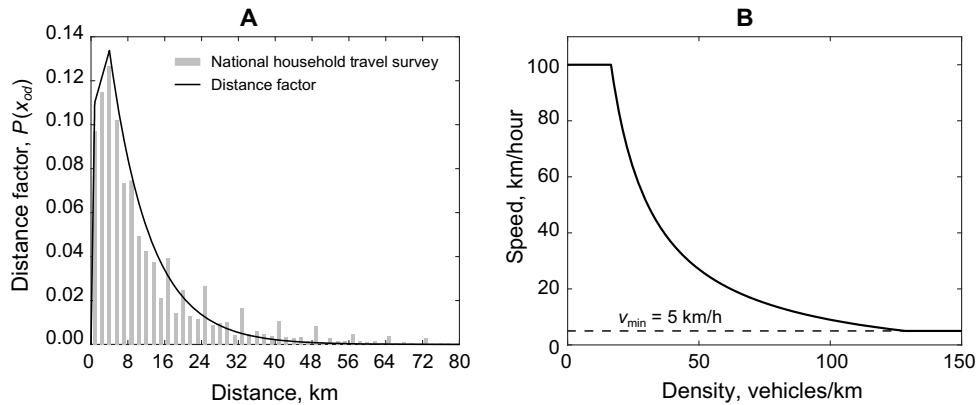

**Fig. 2. Model details.** (**A**) Distance factor $P(x_{od})$ (Eq. 2) of trips given the distance between nodes (solid line) and the statistical data (bars). (**B**) Dependency of speed on density for $V = 100$ km/hour.

We assumed that (i) the flow of commuters from origin region $o$ to destination region $d$ is proportional to the population at the destination $N_d$ and that (ii) the flow of commuters depends on the distance $x_{od}$ between the origin and destination and is given by a distance factor, $P(x_{od})$. Using these assumptions, we assessed the fraction of individuals commuting from region $o$ to destination region $d$, $f_{od}$, as

$$f_{od} = \frac{N_d P(x_{od})}{\sum_k N_k P(x_{ok})} \qquad (2)$$

Then, the commuter flow from origin region $o$ to destination region $d$ is

$$F_{od} = N_o f_{od} \qquad (3)$$

Although individual driving habits may vary (54), we assumed that all drivers tended to optimize their commute paths such that their travel time was minimized. This assumption allowed us to calculate commute paths for every origin-destination pair using inferred FFSs. To calculate commuter flows between all pairs of intersections, we estimated distances $x_{od}$ as the distance of the shortest time path from $o$ to $d$. Furthermore, in place of the distance factor $P(x_{od})$, we used the distribution of trip lengths from the U.S. Federal Highway Administration National Household Travel Survey (55, 56), which we approximated with the exponential function (Fig. 2A and table S3).

Next, we defined the commuter load on each road segment as

$$L_{ij} = \sum_{o,d} F_{od} \theta_{od}(ij) \qquad (4)$$

where $\theta_{od}(ij)$ is a binary variable equal to 0 when the link $ij$ is not on the shortest path connecting nodes $o$ and $d$, and 1 otherwise. Note that in Eq. 4, we only considered origins that were not farther than 30 km from the urban area boundary polygon. The nodes farther than 30 km from the boundary were only used as destinations to evaluate the fraction of commuters not going toward the urban area (Eq. 2).

Because most commuters travel during peak periods, commuter loads $L_{ij}$ can be regarded as traffic-based centrality measures estimating the number of individuals using corresponding road segments. Then, the cumulative time lost by all commuters is

$$\Delta T = \beta \sum_{ij \in E} L_{ij} \left( \frac{(l_{ij} + l_0)}{v_{ij}} - \frac{(l_{ij} + l_0)}{V_{ij}} \right) \qquad (5)$$

where $V_{ij}$ and $v_{ij}$ are, respectively, the FFS and the actual traffic speed along the $ij$ road segment, $l_{ij}$ is its length, $l_0$ is the length correction due to traffic signals, and $\beta$ is the proportionality coefficient same for all urban areas. The summation in Eq. 5 includes only links, whose origins and destinations are within the boundary polygon. A similar equation was obtained for the moving delay in the study of Jiang and Adeli (45), where the authors looked at the delay induced from road repairs.

The actual traffic speed $v_{ij}$ depends on many factors including the speed limit, the number of drivers on the road, and road conditions. Although there exist a number of approaches to estimate actual traffic speed (57, 58), we chose to use the Daganzo model (59) to derive the traffic speed, as shown in the Supplementary Materials

$$v_{ij} = \alpha \frac{l_{ij} m_{ij}}{L_{ij}} - v_{\text{veh}}, \text{ subject to } v_{ij} \in [v_{\min}, V_{ij}] \qquad (6)$$

where $v_{\min}$ is the minimum speed in the traffic, $v_{\text{veh}}$ is the correction for the finite size of the car, and $\alpha$ is the proportionality coefficient (Fig. 2B).

## Efficiency and resilience metrics

We measured efficiency as the average annual delay per peak-period auto commuter. In practice, lower delay means higher efficiency. There are multiple ways to map from delays to efficiency, such as taking the inverse values of delays, taking negative values of delays, etc. To avoid ambiguity and facilitate the interpretation of results, we used the delays themselves to quantify the transportation efficiency of urban areas.

We operationalized resilience through the change in traffic delays relative to stress, which is modeled as loss or impairment of roadway linkages. Looking at resilience from the network science perspective, we focused on topological features of cities, rather than on recovery resources available. Sterbenz *et al.* (60) evaluated a network's resilience as a range of operational conditions for which it stays in the acceptable service region and highlighted that remediation mechanisms drive the operational state toward improvement. We are studying how availability of alternate routes helps remediate the consequences of the initial disruption to the network. In the traffic context, the immediate impact of a given physical disruption (and the time for it to unfold) in terms of closing lanes or reducing speed limits on affected roads will not vary much from network to network, although the number and type of these disruptions will. Likewise, the speed of restoring full functionality (through action in the physical domain) is not so much dependent on the road network as it is on the nature of the disruption (snow versus earthquake versus flood) and the resources that the city allocates to such repair. The level of functionality that these repairs achieve ought to be the full predisruption functionality, that is, eventually all roads can be fully cleared or restored. However, the immediate loss of function for a given traffic flow can very quickly be partially recovered after a disruption by action in the information domain, namely, rerouting of traffic. From the new steady state at that level of functionality, full functionality is gradually restored. Thus, our model proxies for resilience and is calibrated against the data that proxy for efficiency. At the same time, we note that to fully capture resilience characteristics of a transportation system, it is required to analyze recovery resources available and the effectiveness of coordination between the relevant authorities. Lower additional delay corresponds to higher resilience, but using the same reasoning that we had for efficiency, we quantified resilience through additional delays.

## RESULTS

### Efficiency

Together, our traffic model has three parameters (proportionality coefficient $\alpha$, minimum speed $v_{min}$, and finite vehicle size correction $v_{veh}$) and is summarized in Eqs. 5 and 6. Given parameter values of the model, one can estimate the total delay incurred by all commuters in any given suburban area or, equivalently, the average delay per commuter. We take $v_{veh} = 9$ km/hour and $v_{min} = 5$ km/hour and calibrate the model to determine the value of $\alpha$ to match the real data on the annual average delay per peak-period auto commuter provided by the Urban Mobility Scorecard (11).

We divide the 40 urban areas into two equally sized groups for model calibration and validation, respectively. We have found that for the 20 urban areas used for calibration, the $R$-squared coefficient took values in the range (−0.01 to 0.83) (Fig. 3 and Supplementary Materials). This allows us to set model parameters $\alpha$ and $\beta$ (see Methods) as follows: $\alpha = 4.30 \times 10^4$ hour$^{-1}$ and $\beta = 10.59$. These values correspond to the Pearson coefficient of 0.91 ($P = 2.17 \times 10^{-8}$).

To validate the model, we estimate travel delays in 20 different urban areas. As seen from Fig. 3, the estimated travel delays are significantly correlated ($R = 0.63$, $P = 3.00 \times 10^{-3}$) with actual delay times [11], validating the transportation model. Figure 4 is a Google Maps representation of real and modeled results for Los Angeles and San Francisco. Road conditions under real, average traffic patterns at 8 a.m. provided by Google Maps are in Fig. 4 (A and D). Modeled conditions are given for comparison in Fig. 4 (B and E). Finally, Fig. 4 (C and F) shows the new, modeled traffic patterns that result from redistribution of travel in response to a disruption of 5% of the links.

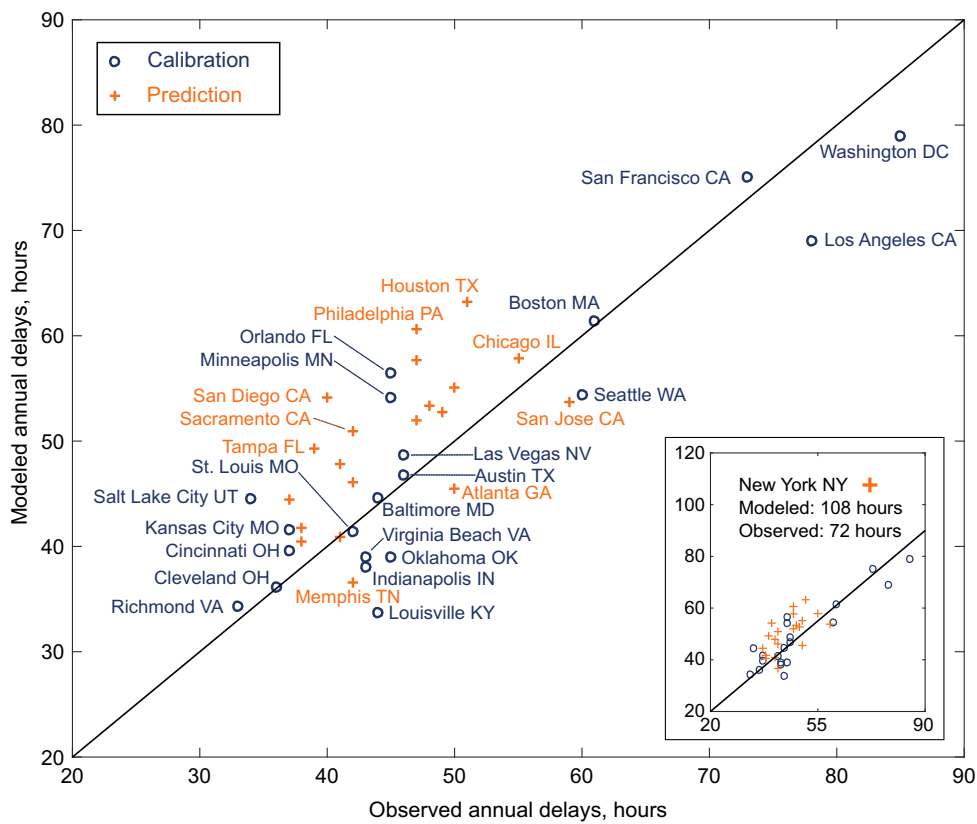

**Fig. 3. Modeled and observed delays in 40 urban areas.** Pearson correlation coefficients and $P$ values between observed and modeled delays are (0.91, 2.17 × 10$^{-8}$) for the 20 cities used to calibrate the model and (0.63, 3.00 × 10$^{-3}$) for the 20 cities used to validate the model. Observed delays were taken from the Texas A&M Transportation Institute Urban Mobility Scorecard (11).

## Resilience

Our approach to model stress is inspired by percolation theory. For every independent simulation of stress, we select a finite fraction of affected road segments $r$ at random, with the probability of failure proportional to segment length. We collect statistics for 20 realizations of the percolation. On failed segments, free-flow speeds (FFSs) are reduced to 1 km/hour (representing near-total loss), and loads $L$ and traffic delays are then recalculated using the updated FFSs. Low-stress scenarios ($r < 0.1$) might be caused by accidents or construction. Larger disruptions might occur during power failures that disrupt traffic signals or severe flooding that makes many roadways nearly impassable. Finally, widespread stress might be caused by snow, ice, or dust storms that affect nearly the entire roadway system. Figure 5 displays the analysis of delay times in six representative urban areas for the full spectrum of adverse event severities, $r \in [0; 1]$. In addition, fig. S5 shows the results for all urban areas. Some routes within a single urban area experience longer delays than others. The inset of Fig. 5 shows the delay distribution for both Los Angeles, which is narrowly clustered, and Boston, where greater variability between roadways is evident.

Traffic delay times grow rapidly as $r$ increases and reach saturation (all routes moving at 1 km/hour) as $r$ approaches 1. We determine the most resilient urban transportation network to be Salt Lake City, UT, whereas the least resilient among the 40 metropolitans is shown to be Washington, DC.

Figure 6 shows both the efficiency (in blue) and resilience response (additional delays due to 5% link disruption, in orange) for the 40 urban areas modeled. Some cities with high efficiency under normal operating conditions (that is, low delays) nevertheless exhibit low resilience (that is, a sharp increase in traffic delays) under stress. Virginia Beach, VA; Providence, RI; and Jacksonville, FL all fall into this category of urban areas in which traffic operates well under ordinary circumstances but rapidly become snarled under mild stress. On the other hand, Los Angeles is notorious for traffic delays under all conditions—yet minor stress levels result in little degradation of efficiency. By contrast, normal traffic delays in San Francisco are comparable to Los Angeles, but mild stress in San Francisco results in large increases in additional delays. These examples indicate that resilience (that is, additional delay response to stress) is independent of normal operating efficiency.

## DISCUSSION

The disturbances affecting the road infrastructure are often complex, and their impact on the structure and function of roadway systems may be unknown [28, 31]. These disturbances might be natural and irregular, such as distributed road closures caused by an earthquake or homogeneous vehicle slowing down because of a snowstorm. The disturbances might also be anthropogenic and intentional, such as a

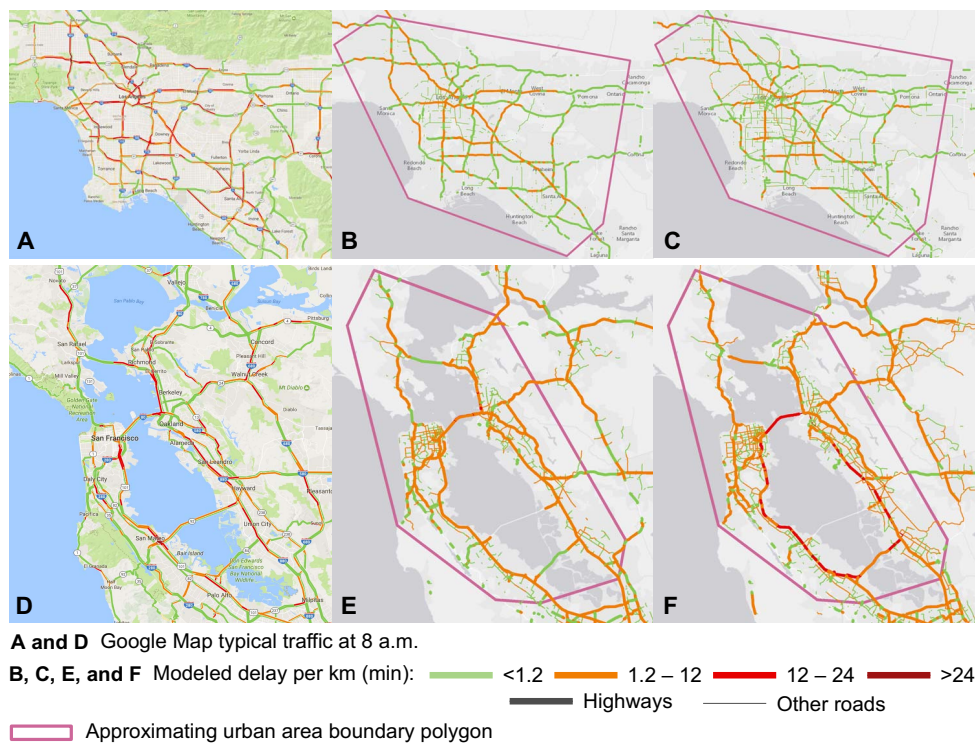

**A and D** Google Map typical traffic at 8 a.m.

**B, C, E, and F** Modeled delay per km (min): ▬ <1.2   ▬ 1.2 – 12   ▬ 12 – 24   ▬ >24   ▬ Highways   ▬ Other roads

▭ Approximating urban area boundary polygon street fair or marathon race. Whatever the disturbance, the results of this analysis allow several meaningful inferences to be made that may have important implications for highway transportation policy. The first is that resilience and efficiency represent different aspects related to the nature of transportation systems; they are not correlated and should be considered jointly as complementary characteristics of roadway networks.

**Fig. 4. Traffic distributions.** Typical congestion at 8 a.m. for Los Angeles (top) and San Francisco (bottom) as given by Google Maps (**A** and **D**), modeled with no disruptions (**B** and **E**), and modeled with a 5% link disruption (**C** and **F**). Notably, in Los Angeles, the disruption results in traffic redistribution to smaller roads, whereas in San Francisco, it results in increased congestion along the major highways.

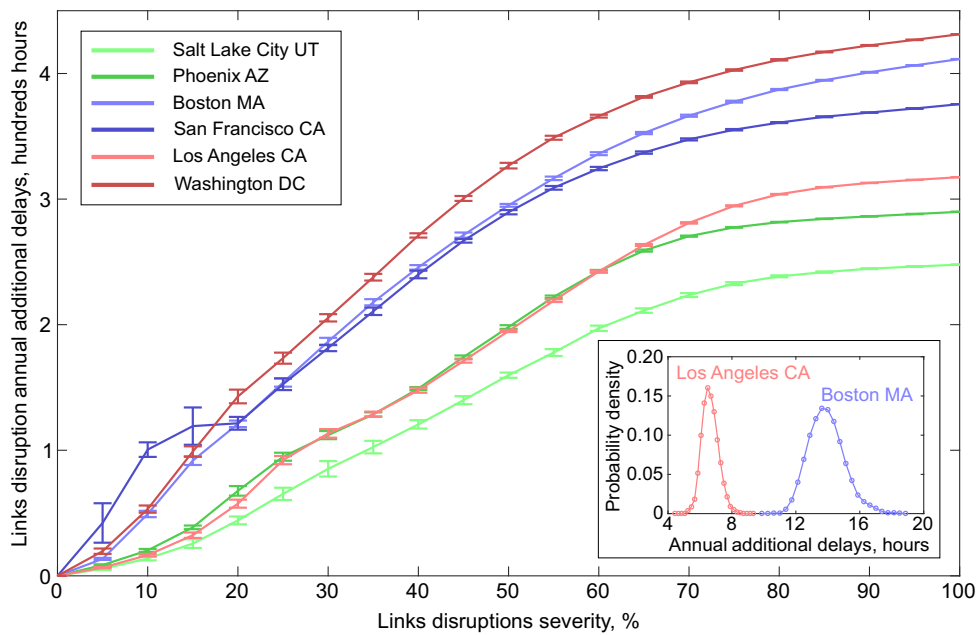

**Fig. 5. Dependency of the additional delay on the severity of the links disruption for six representative urban areas.** Error bars show mean values ± SD. The inset shows distribution densities for two selected urban areas for 1000 realizations of 5% disruption. Note that San Francisco's unique topology makes it susceptible to failures of a small number of discrete roadways, and this produces an anomalous impact at 5 to 15% disruption.

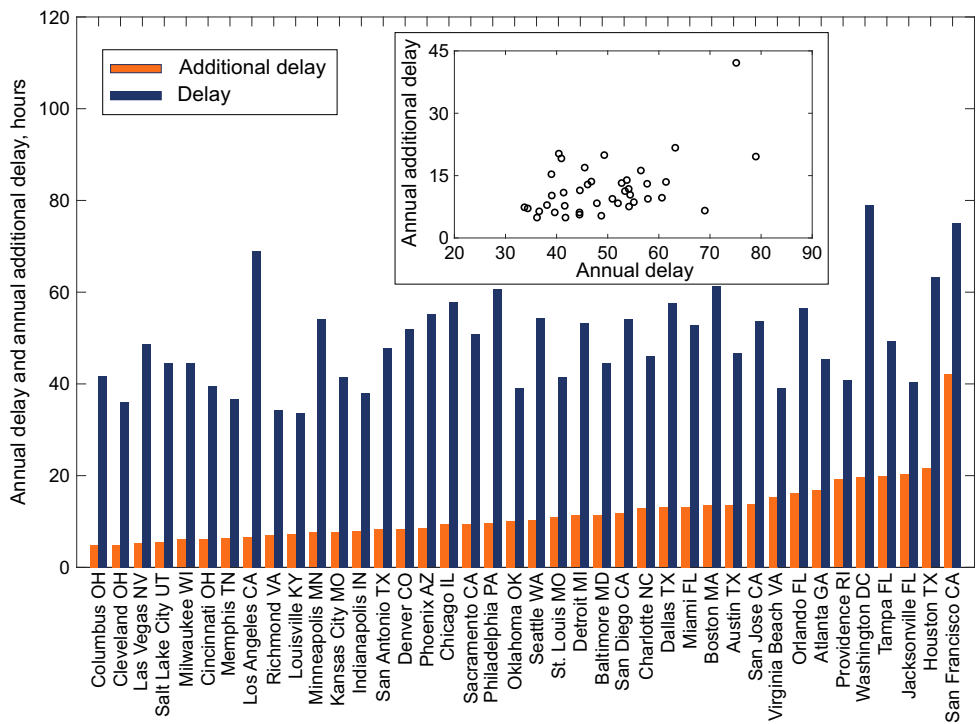

**Fig. 6. Comparison of resilience and efficiency metrics.** Annual impact of 5% disruption (additional delay) has a low correlation with normal annual delay per peak-period auto commuter (delay). Pearson $R = 0.49$, $P = 1.18 \times 10^{-3}$.

Second, there are characteristic differences in the resilience of different urban areas, and these differences are persistent at mild, medium, or widespread levels of stress (Fig. 5). Except for San Francisco,

CA, which is the most fragile of all cities represented in Fig. 5 at stress levels $r < 20\%$ but then surpassed by Boston, MA and Washington, DC, the rank ordering of urban area resilience is insensitive to stress levels. That is, cities that exhibit relatively low resilience under mild stress are the same cities that exhibit low levels of resilience (relative to peers) under widespread roadway impairment. This suggests that the characteristics that impart resilience (such as availability or alternate routes through redundancy of links) are protective against both the intermittent outages caused by occasional car crashes and those caused by snow and ice storms. For cities without resilience, a widespread hazard such as snow may lead to a cascade of conditions (for example, crashes) that rapidly deteriorate into gridlock. This suggests the case for Washington, DC 20 January 2016 under only $2.5 \times 10^{-2}$ m or 2.5 cm of snow (61), and for Atlanta, GA 2 years earlier, which experienced $5.1 \times 10^{-2}$ m or 5.1 cm of snow in the middle of the day that resulted in traffic jams that took days to disentangle (62). Whereas popular explanations of these traffic catastrophes focus on the failure of roadway managers to prepare plows and emergency response equipment, Fig. 5 suggests that cities with similar climates (Memphis, TN and Richmond, VA) are less likely to be affected, regardless of the availability of plow or sand trucks.

The third inference follows from Fig. 6, which suggests that urban areas that make capital investments to reduce traffic delays under normal operating conditions may nevertheless be vulnerable to traffic delays under mild stress conditions. Because these stressors are inevitable, whether from crashes, construction, special events, extreme weather, equipment malfunctions, or even deliberate attack, investment strategies that prioritize reduction of normal operating delays may have the unintended consequence of exacerbating tail risks—that is, the risk of worse catastrophe under unlikely but possible conditions.

Finally, the exceptional position of New York City in Fig. 3 calls attention to the fact that substitutes for roadway transportation are available in many cities and have an important role to play in relieving traffic congestion. According to the Texas A&M Institute (63, 64), public transit reduces delays per peak-period auto commuter in the New York urban area by 63 hours, in Chicago by 23 hours, and by less than 20 hours in other urban areas. Because our model considers only roadway transit, and New York City contains a myriad of nonroad-based options to avoid roadway congestion, it is unlikely that our model can provide informative results for the New York urban area.

Although interest has increased in policies that enhance roadway resilience, few analytic tools are available to guide new investments in achieving resilience goals. It is well understood that roadway infrastructure is expensive, both in acquiring land for rights-of-way and in construction of improvements, and thus, decisions regarding alignment, crossing, and access made over a period of decades may have long-lasting consequences that are observable in traffic data today. Consequently, urban areas exhibit different unintentional traffic characteristics, including delays under normal and random stress conditions. Investments motivated exclusively by expected efficiencies under normal operating conditions are unreliable safeguards against loss of efficiency under stress conditions. Therefore, new analytic tools are required that allow designers to assess the adaptive capacity of roadway infrastructure and assess the potential of new investments to provide enhanced resilience. The adaptive network-based model described herein is one such approach.

## SUPPLEMENTARY MATERIALS

Supplementary material for this article is available at http://advances.sciencemag.org/cgi/content/full/3/12/e1701079/DC1

Alternative approaches to model transportation

Mapping from OSM Foundation shapefiles to network nodes and links

Population assignment algorithm

Distance factor of the likelihood of travel between nodes

Estimation of the traffic speed from the density of vehicles

Model calibration procedure

Sensitivity of the model to ramp speeds

Additional delay as a function of the severity of link disruption table S1. Mapping original OSM types to network link types and assignment of the number of lanes.

table S2. The algorithm of the node population assignment.

table S3. Distance factor $P(x_{od})$ of the likelihood of travel between nodes.

table S4. Model sensitivity to ramp speed coefficient.

fig. S1. Effects of the removal of nodes of degree 2.

fig. S2. Density-flow relationship in the Daganzo traffic model.

fig. S3. Model calibration.

fig. S4. Modeled delays for ramp speed coefficients of $1/3$ and $1/2$.

fig. S5. Dependency of the additional delay on the severity of the link disruption for all 40 urban areas.

## REFERENCES AND NOTES

1. A. Samuelsson, B. Tilanus, A framework efficiency model for goods transportation, with an application to regional less-than-truckload distribution. *Transp. Logist.* **1**, 139–151 (1997).
2. K. Beverly, *Efficient Use of Highway Capacity* (FHWA-HOP-10-023, Texas Transportation Institute, 2010), p. 100.
3. R. Hoogendoorn, B. van Arem, S. Hoogendoorn, Automated driving, traffic flow efficiency, and human factors: Literature review. *Transp. Res. Rec.* **2422**, 113–120 (2014).
4. J. Sami, F. Pascal, B. Younes, Public road transport efficiency: A stochastic frontier analysis. *J. Transp. Syst. Eng. Inf. Technol.* **13**, 64–71 (2013).
5. V. Sanchez Rodrigues, D. Stantchev, A. Potter, M. Naim, A. Whiteing, Establishing a transport operation focused uncertainty model for the supply chain. *Int. J. Phys. Distrib. Logist. Manag.* **38**, 388–411 (2008).
6. S. E. Chang, N. Nojima, Measuring post-disaster transportation system performance: The 1995 Kobe earthquake in comparative perspective. *Transp. Res. A Policy Pract.* **35**, 475–494 (2001).
7. G. M. D'Este, R. Zito, M. A. P. Taylor, Using GPS to measure traffic system performance. *Comput. Aided Civ. Inf. Eng.* **14**, 255–265 (1999).
8. G. Yan, T. Zhou, B. Hu, Z.-Q. Fu, B.-H. Wang, Efficient routing on complex networks. *Phys. Rev. E* **73**, 046108 (2006).
9. W. B. Allen, D. Liu, S. Singer, Accessibility measures of US metropolitan areas. *Transp. Res. B Methodol.* **27**, 439–449 (1993).
10. T. Yamashita, K. Izumi, K. Kurumatani, Car navigation with route information sharing for improvement of traffic efficiency, in *Proceedings of the 7th International IEEE Conference on Intelligent Transportation Systems* (IEEE, 2004), pp. 465–470.
11. D. Schrank, B. Eisele, T. Lomax, J. Bak, *2015 Urban Mobility Scorecard* (Texas A&M Transportation Institute and INRIX, 2015); http://d2dtl5nnlpfr0r.cloudfront.net/tti.tamu.edu/documents/mobility-scorecard-2015.pdf.
12. S. Çolak, A. Lima, M. C. González, Understanding congested travel in urban areas. *Nat. Commun.* **7**, 10793 (2016).
13. K. Turnbull, Technical Activities Division, Transportation Research Board, National Academies of Sciences, Engineering, and Medicine, *Transportation Resilience: Adaptation to Climate Change* (Transportation Research Board, 2016).
14. C. S. Holling, Resilience and stability of ecological systems. *Annu. Rev. Ecol. Syst.* **4**, 1–23 (1973).
15. C. S. Holling, Engineering Resilience versus Ecological Resilience, in *Engineering within Ecological Constraints*, P. C. Schulze, Ed. (The National Academies Press, 1996), pp. 31–44.
16. I. Linkov, T. Bridges, F. Creutzig, J. Decker, C. Fox-Lent, W. Kröger, J. H. Lambert, A. Levermann, B. Montreuil, J. Nathwani, R. Nyer, O. Renn, B. Scharte, A. Scheffler, M. Schreurs, T. Thiel-Clemen, Changing the resilience paradigm. *Nat. Clim. Change* **4**, 407–409 (2014).
17. A. A. Ganin, E. Massaro, A. Gutfraind, N. Steen, J. M. Keisler, A. Kott, R. Mangoubi, I. Linkov, Operational resilience: Concepts, design and analysis. *Sci. Rep.* **6**, 19540 (2016).
18. J. Park, T. P. Seager, P. S. C. Rao, Lessons in risk- versus resilience-based design and management. *Integr. Environ. Assess. Manag.* **7**, 396–399 (2011).
19. S. E. Flynn, S. P. Burke, *Critical Transportation Infrastructure and Societal Resilience* (Center for National Policy, 2012).

20. D. Marchese, I. Linkov, Can you be smart and resilient at the same time? *Environ. Sci. Technol.* **51**, 5867–5868 (2017).

21. D. L. Alderson, G. G. Brown, W. M. Carlyle, Operational Models of Infrastructure Resilience. *Risk Anal.* **35**, 562–586 (2015).

22. D. D. Woods, Four concepts for resilience and the implications for the future of resilience engineering. *Reliab. Eng. Syst. Safe.* **141**, 5–9 (2015).

23. T. P. Seager, S. Spierre Clark, D. A. Eisenberg, J. E. Thomas, M. M. Hinrichs, R. Kofron, C. N. Jensen, L. R. McBurnett, M. Snell, D. L. Alderson, Redesigning resilient infrastructure research, in *Resilience and Risk*, I. Linkov, J. M. Palma-Oliveira, Eds. (Springer, 2017).

24. D. Freckleton, K. Heaslip, W. Louisell, J. Collura, Evaluation of transportation network resiliency with consideration for disaster magnitude, paper presented at the 91st Annual Meeting of the Transportation Research Board, Washington, DC, 2012.

25. S. B. Pant, Transportation network resiliency: A study of self-annealing, thesis, Utah State University, Logan (2012).

26. D. King, A. Shalaby, Performance metrics and analysis of transit network resilience in Toronto, paper presented at the 95th Annual Meeting of the Transportation Research Board, Washington, DC, 10 to 14 January 2016.

27. D. Li, Resilience of spatial networks, in *Complex Systems and Networks*, J. Lü, X. Yu, G. Chen, W. Yu, Eds. (Springer Berlin Heidelberg, 2016), pp. 79–106.

28. A. Kermanshah, S. Derrible, Robustness of road systems to extreme flooding: Using elements of GIS, travel demand, and network science. *Nat. Hazards* **86**, 151–164 (2017).

29. A. Cox, F. Prager, A. Rose, Transportation security and the role of resilience: A foundation for operational metrics. *Transp. Policy* **18**, 307–317 (2011).

30. M. Carey, C. Hendrickson, Bounds on expected performance of networks with links subject to failure. *Networks* **14**, 439–456 (1984).

31. P. M. Murray-Tuite, A Comparison of Transportation Network Resilience under Simulated System Optimum and User Equilibrium Conditions, in *Proceedings of the Winter Simulation Conference WSC 06*, 3 to 6 December 2006, pp. 1398–1405.

32. S. Melkote, M. S. Daskin, An integrated model of facility location and transportation network design. *Transp. Res. A Policy Pract.* **35**, 515–538 (2001).

33. A. Thiagarajan, L. Ravindranath, K. LaCurts, S. Madden, H. Balakrishnan, S. Toledo, J. Eriksson, VTrack: Accurate, energy-aware road traffic delay estimation using mobile phones, in *Proceedings of the 7th ACM Conference on Embedded Networked Sensor Systems* (ACM Press, 2009), p. 85.

34. S. Simonovic, From risk management to quantitative disaster resilience – a paradigm shift. *Int. J. Saf. Secur. Eng.* **6**, 85–95 (2016).

35. 2012 Cartographic Boundary File, 2010 Census Urban Area for United States, 1:500,000 (U.S. Census Bureau, 2012); www2.census.gov/geo/tiger/TIGER2010/UA/2010/tl_2010_us_uac10.zip.

36. 2010 Census Data (U.S. Census Bureau, 2010); ftp://ftp2.census.gov/geo/tiger/TIGER2010DP1/Tract_2010Census_DP1.zip.

37. Planet data (OpenStreetMap contributors, 2017); https://planet.openstreetmap.org.

38. Y. Berezin, A. Bashan, M. M. Danziger, D. Li, S. Havlin, Localized attacks on spatially embedded networks with dependencies. *Sci. Rep.* **5**, 8934 (2015).

39. A. Majdandzic, L. A. Braunstein, C. Curme, I. Vodenska, S. Levy-Carciente, H. E. Stanley, S. Havlin, Multiple tipping points and optimal repairing in interacting networks. *Nat. Commun.* **7**, 10850 (2016).

40. J. Zhao, D. Li, H. Sanhedrai, R. Cohen, S. Havlin, Spatio-temporal propagation of cascading overload failures in spatially embedded networks. *Nat. Commun.* **7**, 10094 (2016).

41. M. C. González, C. A. Hidalgo, A.-L. Barabási, Understanding individual human mobility patterns. *Nature* **453**, 779–782 (2008).

42. E. Cho, S. A. Myers, J. Leskovec, Friendship and mobility: User movement in location-based social networks, in *Proceedings of the 17th ACM SIGKDD International Conference on Knowledge Discovery and Data Mining* (ACM Press, 2011), p. 1082.

43. F. Calabrese, M. Diao, G. Di Lorenzo, J. Ferreira Jr., C. Ratti, Understanding individual mobility patterns from urban sensing data: A mobile phone trace example. *Transp. Res. C Emerg. Technol.* **26**, 301–313 (2013).

44. D. Gross, J. F. Shortle, J. M. Thompson, C. M. Harris, *Fundamentals of Queueing Theory* (Wiley Series in Probability and Statistics, Wiley, Hoboken, NJ, ed. 4, 2008).

45. X. Jiang, H. Adeli, Freeway work zone traffic delay and cost optimization model. *J. Trans. Eng.* **129**, 230–241 (2003).

46. Growth in Urban Population Outpaces Rest of Nation, Census Bureau Reports (U.S. Census Bureau, 2012); www.census.gov/newsroom/releases/archives/2010_census/cb12-50.html.

47. H. J. Casey, Applications to traffic engineering of the law of retail gravitation. *Traffic Q.* **IX** (1955), pp. 23–35.

48. M. Sabyasachee, Y. Wang, X. Zhu, R. Moeckel, S. Mahapatra, Comparison between gravity and destination choice models for trip distribution in Maryland, paper presented at the TRB 92nd Annual Meeting of Compendium of Papers, 13 to 17 January 2013.

49. E. Cascetta, F. Pagliara, A. Papola, Alternative approaches to trip distribution modelling: A retrospective review and suggestions for combining different approaches. *Pap. Reg. Sci.* **86**, 597–620 (2007).

50. J. de Dios Ortúzar, L. G. Willumsen, *Modelling Transport* (John Wiley & Sons, ed. 4, 2011).

51. National Research Council (U.S.), *Metropolitan Travel Forecasting: Current Practice and Future Direction* (Transportation Research Board, 2007).

52. A. S. Fotheringham, Some theoretical aspects of destination choice and their relevance to production-constrained gravity models. *Environ. Plan.* **15**, 1121–1132 (1983).

53. A. G. Woodside, S. Lysonski, A general model of traveler destination choice. *J. Travel Res.* **27**, 8–14 (1989).

54. A. Lima, R. Stanojevic, D. Papagiannaki, P. Rodriguez, M. C. González, Understanding individual routing behaviour. *J. R. Soc. Interface* **13**, 20160021 (2016).

55. 2009 National Household Travel Survey (U.S. Department of Transportation, Federal Highway Administration, 2009); http://nhts.ornl.gov.

56. R. Van Haaren, *Assessment of Electric Cars' Range Requirements and Usage Patterns based on Driving Behavior recorded in the National Household Travel Survey of 2009* (Solar Journey, 2012), p. 25.

57. B. D. Greenshields, J. R. Bibbins, W. S. Channing, H. H. Miller, R. W. Crum, A study of traffic capacity, in *Proceedings of the 14th Annual Meeting of the Highway Research Board*, 6 to 7 December 1934, vol. 1.

58. S. Smulders, Control of freeway traffic flow by variable speed signs. *Transp. Res. B Methodol.* **24**, 111–132 (1990).

59. C. F. Daganzo, The cell transmission model: A dynamic representation of highway traffic consistent with the hydrodynamic theory. *Transp. Res. B Methodol.* **28**, 269–287 (1994).

60. J. P. G. Sterbenz, D. Hutchison, E. K. Çetinkaya, A. Jabbar, J. P. Rohrer, M. Schöller, P. Smith, Resilience and survivability in communication networks: Strategies, principles, and survey of disciplines. *Comput. Netw.* **54**, 1245–1265 (2010).

61. T. Richardson, *Evening snow brings traffic gridlock across Washington* (The Washington Post, 2016); www.washingtonpost.com/news/dr-gridlock/wp/2016/01/20/evening-snow-brings-traffic-gridlock-across-washington/?utm_term=.6aada266526a.

62. C. Sen, *How 2 Inches of Snow Created a Traffic Nightmare in Atlanta* (The Atlantic, 2014); www.theatlantic.com/business/archive/2014/01/how-2-inches-of-snow-created-a-traffic-nightmare-in-atlanta/283434/.

63. A. M. Renn, *The Cost of Congestion, The Value of Transit* (Urbanophile LLC, 2011); www.urbanophile.com/2011/10/06/the-cost-of-congestion-the-value-of-transit/.

64. T. Lomax, D. Schrank, S. Turner, L. Geng, Y. Li, N. Koncz, *Real-Timing the 2010 Urban Mobility Report* (UTCM 10-65-55, Texas Transportation Institute, 2011); http://utcm.tamu.edu/publications/final_reports/Lomax_10-65-55.pdf).

**Acknowledgments:** We would like to thank S. Buldyrev (Yeshiva University) and J. Palma-Oliveira (University of Lisbon) for their insightful comments. **Funding:** This study was supported by the U.S. Army Engineer Research and Development Center and by the Defense Threat Reduction Agency, Basic Research Program (P. Tandy, program manager). A.A.G. was additionally supported by the Virginia Transportation Research Council and Virginia Department of Transportation. T.S. was supported by the NSF under grant no. 1441352. **Author contributions:** A.A.G., M.K., and I.L. conceived the model and designed the simulations. A.A.G. developed software and performed data retrieval and simulations. A.A.G. and M.K. analyzed results. I.L. provided senior guidance. A.A.G., M.K., J.M.K., T.S., and I.L. wrote the paper and contributed to the interpretation of the results. **Competing interests:** The authors declare that they have no competing interests. **Data and materials availability:** All data needed to evaluate the conclusions in the paper are present in the paper and/or the Supplementary Materials. Additional data related to this paper may be requested from the authors. Map data were copyrighted by OSM contributors and are available at www.openstreetmap.org.

Submitted 6 April 2017
Accepted 21 November 2017
Published 20 December 2017
10.1126/sciadv.1701079

**Citation:** A. A. Ganin, M. Kitsak, D. Marchese, J. M. Keisler, T. Seager, I. Linkov, Resilience and efficiency in transportation networks. *Sci. Adv.* **3**, e1701079 (2017).

# Science Advances

## Resilience and efficiency in transportation networks

Alexander A. Ganin, Maksim Kitsak, Dayton Marchese, Jeffrey M. Keisler, Thomas Seager and Igor Linkov

| | |
|---|---|
| **ARTICLE TOOLS** | http://advances.sciencemag.org/content/3/12/e1701079 |
| **SUPPLEMENTARY MATERIALS** | http://advances.sciencemag.org/content/suppl/2017/12/18/3.12.e1701079.DC1 |
| **REFERENCES** | This article cites 35 articles, 1 of which you can access for free http://advances.sciencemag.org/content/3/12/e1701079#BIBL |
| **PERMISSIONS** | http://www.sciencemag.org/help/reprints-and-permissions |

Use of this article is subject to the Terms of Service